\documentclass[twocolumn,showpacs,preprintnumbers,amsmath,amssymb,pra]{revtex4}
\usepackage{epsfig}
\usepackage{epstopdf}

\begin{document}

\title{Many-particle Systems in One Dimension in the Harmonic Approximation}
\author{Armstrong J R, Zinner N T, Fedorov D V and Jensen A S}
\affiliation{Department of Physics and Astronomy, Aarhus University, 
DK-8000 Aarhus C, Denmark}

\date{\today}
\begin{abstract}
We consider energetics and structural properties of a many particle system in one
dimension with pairwise contact interactions confined in a parabolic external 
potential. To render the problem analytically solvable, we use the 
harmonic approximation scheme at the level of the Hamiltonian. We investigate
the scaling with particle number of the ground state energies for systems consisting
of identical bosons or fermions. We then proceed to focus on bosonic systems and 
make a detailed comparison to known exact results in the absence of the parabolic 
external trap for three-body systems. We also consider the thermodynamics of the 
harmonic model which turns out to be similar for bosons and fermions due to the 
lack of degeneracy in one dimension.
\end{abstract}

\maketitle
\section{Introduction}
Exactly solvable models are key players in few- and many-body quantum mechanics and are indeed also 
fascinating creatures \cite{mattis,sutherland}. The analytical intractability
of general $N$-body systems makes it extremely important to have exact results to allow 
benchmark tests of complicated numerical methods. Unfortunately, exactly solvable models are 
few and far between and are more often found in low-dimensional systems. In the case of 
one spatial dimension, the famous Bethe ansatz \cite{bethe1931} was succesfully 
applied to models of bosons with zero-range interactions \cite{lieb1963a,lieb1963b,macguire1964}, 
and later to interactions with long range \cite{sutherland1971,calogero1971}.

Meanwhile, an exciting direction in the field of 
cold atomic gases aims at the study of 
low-dimensional systems in general and one-dimensional setups in particular \cite{bloch2008}. 
A noticable highlight of this pursuit is the experimental realization of the so-called Tonks-Girardeau
gas \cite{paredes2004,kinoshita2004,haller2009}, where strongly-interacting one-dimensional 
bosons become impenetrable objects and behave similar to fermions \cite{tonks1936,girardeau1960}.
Very recently, it has even become possible to study this interesting regime in the limit of 
small particle number \cite{serwane2011,zurn2012}.

Here we study an exactly solvable model of an $N$-body system in an external parabolic
confinement. The outer trap is always present in cold atomic gas experiments but can 
often be neglected or treated in a local density approximation when large systems are 
studied. However, for smaller particle numbers the effect of the outer trap becomes 
important for the structure and dynamics of the system. To make the $N$-body problem
tractable we use a harmonic Hamiltonian approximation \cite{calogero1971} with carefully
chosen parameters that reproduce essential features of two atoms interacting via short-range
interactions in a parabolic trap \cite{arms11}.

The paper is organized as follows. After a discussion of the harmonic methods we 
provide details on how the parameters of the harmonic interactions are obtained
from knowledge of the exact solution of the two-body problem originally obtained
by Busch {\it et al.} \cite{busc98}. We then present results for the ground-state 
energies and the radii. Our focus is on bosonic systems, but we do present a few
results for fermions as well. A comparison to the exact results of MacGuire \cite{macguire1964}
is then made in relevant limits. We also compute the one-body density matrix and its
largest eigenvalue to obtain the condensate fraction at zero temperature. Lastly, 
we discuss the thermodynamics of our model and then proceed to conclusion and outlook 
for future work.

\section{Method}
We consider a system of $N$ quantum particles interacting pairwise via a delta 
function interaction and confined by a harmonic external potential in one-spatial 
dimension.  The Hamiltonian for this system is
\begin{equation}
H=-\frac{\hbar^2}{2m}\sum_{i=1}^N\frac{\partial^2}{\partial x_i^2} +\frac{1}{2}m\omega_0^2\sum_{i=1}^Nx_i^2-\frac{2\hbar^2}{ma}\sum_{i<k}\delta^{1D}(x_i-x_k),
\label{em1}
\end{equation}
where $m$ is the mass of the particles, $\omega_0$ is the frequency of the 
external field, and $a$ is the one-dimensional scattering length which parameterizes the 
strength of the two-body interaction (we will discuss its relation to the three-dimensional 
scattering length below).
The external field, $\omega_0$, defines the length scale of our 
system, $l^2=\hbar/(m\omega_0)$.  For two particles, (\ref{em1}) has 
been solved in, e.g., \cite{busc98,farr10}, and their results of two-body 
energies and wave functions for a given scattering length are used to determine 
the parameters of our model.  We consider only the bound molecular branch 
of the system, i.e., $a>0$.

In our general harmonic approximation scheme we replace Hamiltonian (\ref{em1}) with
\begin{eqnarray}
H&=&-\frac{\hbar^2}{2m}\sum_i^N\frac{\partial^2}{\partial x_i^2}
+\frac{1}{2}m\omega_0^2\sum_i^Nx_i^2\nonumber\\
&& +\frac{1}{2}\mu\omega_{in}^2\sum_{i<k}(x_i-x_k)^2+\sum_{i<k}V_S,
\label{em2}
\end{eqnarray}
where $\mu=m/2$ is the reduced mass of the two-body system, $\omega_{in}$ is 
the interacting frequency, and $V_S$ is an energy shift. The solution 
to this equation for $N$ particles for general systems is described in detail 
in \cite{arms11} and more specifically for identical particles 
in \cite{calogero1971} and \cite{arms12}. 

The 
parameters $\omega_{in}$ and $V_S$ are now chosen to fit pertinent 
properties of the original Hamiltonian at the two-body level in the 
usual spirit of constructive descriptions of $N$-body systems based on 
two-body interactions. Here we impose the condition that at the two-body 
level the harmonic oscillator reproduces the energy and average square 
radius of the exact two-body solution. In order to fulfil these
constraints, we use the size of the system to determine the 
interaction frequency through the relation
\begin{equation}
\frac{\langle\psi|x^2|\psi\rangle}{\langle\psi|\psi\rangle}=\frac{\hbar}{2\mu\sqrt{\omega_{in}^2+\omega_0^2}},
\label{em7}
\end{equation}
where $x=x_1-x_2$ is the relative coordinate in the two-body system. 
The energy shift is determined by requiring that the model reproduces the energy of the two-body system
\begin{equation}
E_2=\frac{1}{2}\hbar\sqrt{\omega_{in}^2+\omega_0^2}+V_S.
\label{em8}
\end{equation}
The quantities $E_2$ and $\langle x^2\rangle$ can be easily obtained by numerically solving the 
transcendental equations fulfilled by the exact solution for two bosons (or two fermions in different
spin states) interacting via a zero-range interaction in a parabolic trap \cite{busc98}.
Note that when we consider identical fermions below, there can in principle only be a 
non-zero two-body interaction in states that are odd under exchange of the two fermions. 
In this paper we are mostly interested in the scaling behaviour with particle number 
of the fermions as the interaction frequency is varied (through the scattering length) and
we therefore ignore this point and use the same input for fermions as for bosons. The 
most important difference is of course the quantum statistics which is taken fully into 
account.
  
We note that in cold atoms, effective one-dimensional setups are created by using a tightly 
confining potential in two transverse directions, usually through the application of an 
optical lattice \cite{bloch2008}. In a deep transverse lattice, the atoms are then effectively only occupying
the lowest transverse motional degree of freedom. This transverse degree of freedom, however,
has an influence on the effective one-dimensional scattering length that described the inter-atomic
interaction in the system. A mapping exists between the true three-dimensional scattering 
length and the effective one-dimensional scattering length that takes the transverse degrees
of freedom explicitly into account \cite{olshanii98}. This mapping should thus be applied
before comparison to realistic experiments.
  
Once the parameters have been determined, (\ref{em2}) can be solved.  
The ground state energy for identical bosons is
\begin{equation}
E_{gs}^{(b)}=\frac{1}{2}N(N-1)V_S+\frac{1}{2}(N-1)\hbar\omega_r+\frac{1}{2}\hbar\omega_0,
\label{em3}
\end{equation}
where 
\begin{equation}
\omega_r^2=N\omega_{in}^2/2+\omega_0^2
\label{em5}
\end{equation}
is the $(N-1)$-degenerate frequency that comes out of the solution of (\ref{em2}) \cite{arms11}.  
The last term in (\ref{em3}) is the energy related to the center of mass motion 
and will be ignored as we are interested in the internal dynamics of the system.

Though most of this work concerns bosons, input for identical fermions is also possible.  
Since fermions obey the Pauli Principle, $N-1$ particles must be placed in higher 
oscillator levels as there are no degeneracies in one dimension.  
We can obtain the amount of energy 
in the ground state by filling 
consecutive oscillator levels, i.e.
\begin{equation}
\hbar\omega_0\sum_{k=0}^{N-1}\left(k+\frac{1}{2}\right)=\frac{N^2}{2}\hbar\omega_0.
\end{equation}
The complete ground state energy is then 
\begin{equation}
E_{gs}^{(f)}=\frac{1}{2}N(N-1)V_S+\frac{N^2}{2}\hbar\omega_r+\frac{1}{2}\hbar\omega_0.
\label{em4}
\end{equation}
This is a different scaling than for the bosons in (\ref{em3}).  In the fermion case, 
the energy is dominated by the $N^2\hbar\omega_r/2$ term which scales as $N^{5/2}$.

\begin{figure}[ht!]
\includegraphics[width=0.5\textwidth]{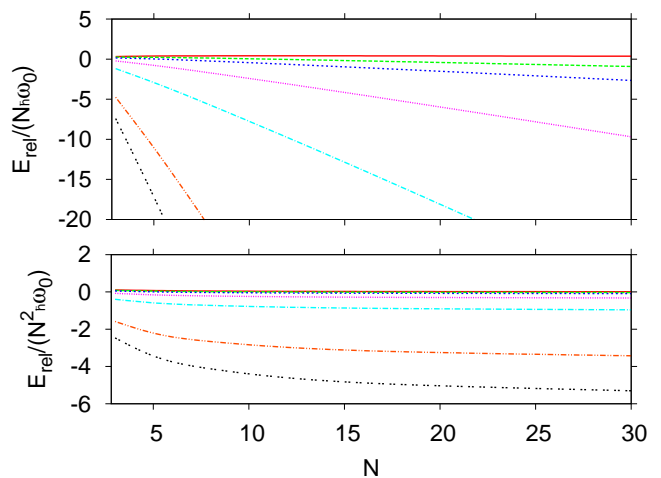}
\caption{  Top panel: Energies per particle for bosons confined on one dimension.  The $y$-axis is stopped at -20 in order to show the general behaviour of many curves, though the shortest scattering lengths quickly leave the figure.  The scattering lengths plotted, moving from bottom to top are 0.4., 0.5, 1.0, 2.0, 5.0, 10.0, and 100.0
Bottom panel: Energies divided by $N^2$ for the bosons to show their scaling behaviour with $N$.  The order of the scattering lengths from bottom to is the same as in the top panel.}
\label{fr1}
\end{figure}

\section{Results}
We now present numerical results for systems with $N=3-30$ particles, studying
their energetics, the radial behaviour of the systems, and the one-body
density matrix. Here we focus on 
the case of identical bosons. A particular issue is the ratio of the three-body
energy to the two-body energy. An exact formula of MacGuire \cite{macguire1964}
applies to one-dimensional systems of bosons with zero-range interactions and
we make a comparison of the harmonic results to that model in the strongly-bound
limit where the two-body energy is large and negative. Lastly, we consider also
the thermodynamics of the system within the harmonic approximation.

\subsection{Energies and radii}
\begin{figure}[ht!]
\includegraphics[width=0.5\textwidth]{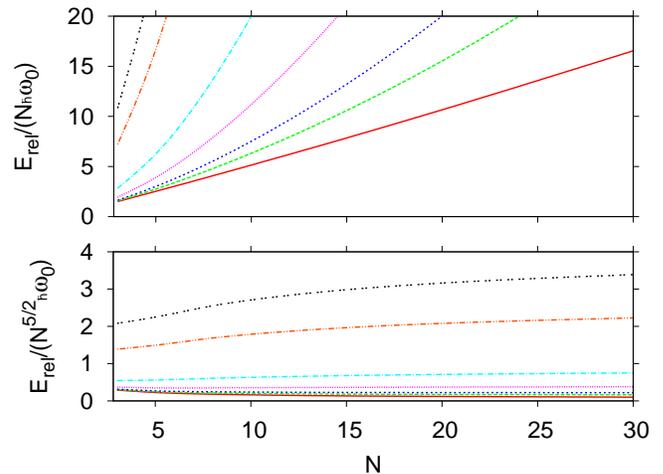}
\caption{Top panel: Energies per particle for fermions confined on one dimension.  
The plotted scattering lengths are in the opposite order from the bosons as moves 
from the bottom to the top of the plot.
Bottom panel: Fermion energies divided $N^{5/2}$ to clearly show the scaling 
of the energy with $N$.}
\label{fr2}
\end{figure}

The ground state energy for bosons is shown in figure \ref{fr1} and that 
for fermions in figure \ref{fr2} for different particle numbers and scattering
legnths. Notice that the upper panels show the energy per particle, while in the 
lower panels the energy is divided by a different power that we will discuss below.
A striking feature to notice is the positivity of the energy for fermions, while 
that of the bosons can have both signs. This is a consequence of the Pauli
principle which implies that the fermions will have to occupy higher 
orbitals for lack of degeneracies in one dimension as discussed above. 
Even in the case of very small scattering lengths (and thus large and negative two-body
binding energies) the contribution from the higher orbits makes the overall
ground state energy positive.

Once the two-body problem is solved and the parameters are determined, the ground 
state energy for bosons is given by (\ref{em3}).  As seen in figure \ref{fr1}, 
the sign of the energy of the Bose system is mostly negative, but for small 
systems at large scattering lengths, the energy turns positive.  
This is understood from the fact that the energy shift, $V_S$, 
is negative and the shift term in the energy of (\ref{em3}) scales with $N^2$ 
while the positive oscillator frequency term scales with the lower power $N^{3/2}$.
For bosons, the energy is positive for all particle numbers examined for $a/l=100$.
Using (\ref{em3}), one can derive the critical number for bosons where the 
energy changes sign. It is given by
\begin{equation}
N_{crit}=\frac{\hbar^2\omega_{in}^2+2\hbar\sqrt{\hbar^2\omega_{in}^4/4+4V_S^2\omega_0^2)}}{4V_S^2}.
\end{equation}  
From this relation one can calculate that for bosons the energy 
becomes negative at $a/l=100$ when $N\geq 116$. 

The bottom panels of Figures \ref{fr1} and \ref{fr2} 
show the energies divided by the relevant scaling factor ($N^2$ for bosons and $N^{5/2}$
for fermions) to emphasize the scaling behaviour at large $N$. In the figures we 
can see that the asymptotic scalings are nicely approached when the scattering length
is not too small. For the case of small scattering length, the two-body energy will 
in general behave as $E_2\sim -1/a^2$ while $\langle x^2\rangle\sim a^2$, 
since in this limit the trap can be ignored
on the bound state branch in the spectrum that we study here. However, the 
relations in (\ref{em7}) and (\ref{em8}) that we use to determine our oscillator 
parameters now imply that both $V_S$ and $\omega_{in}$ must scale with $a^{-2}$
to be fulfilled (remember that $E_2$ is negative). This means that there will 
be a competition between the terms in the ground state energies given by (\ref{em3}) and 
(\ref{em4}). The asymptotic behaviour for large $N$ is therefore approached more slowly
for small $a$.

\begin{figure}[ht!]
\includegraphics[width=0.5\textwidth]{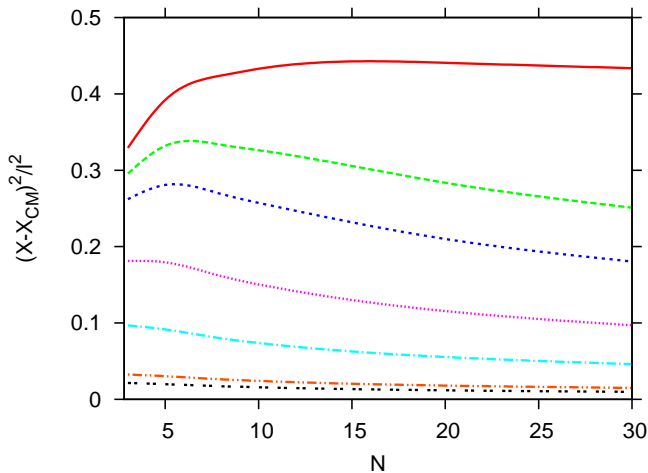}
\caption{  Radius $(x-X_{CM})^2$ for bosons as a function of particle number 
for several different scattering lengths.  The scattering lengths are, 
from bottom to top, 0.4, 0.5, 1, 2, 5, 10, and 100.}
\label{fr5}
\end{figure}

We also calculate the relative size of the bosonic ground state wave 
functions, $\langle \left(X-X_{CM}\right)^2\rangle$, as a function of 
particle number for several different scattering lengths. Note here that 
$X$ denotes the single particle coordinate of one of the bosons and that no
index is needed since the particles are identical.
The results are 
shown in figure \ref{fr5}.  One can see for the large scattering lengths that 
the radius increases before eventually decreasing for larger particle numbers 
(more than 20 for $a/l=100$).  For scattering lengths of two and smaller, the 
size decreases monotonically with increasing particle number. Again this is
connected to the fact that the small $a$ regime has very strong two-body 
binding and thus small $\langle x^2\rangle$ which is imprinted on the $N$-body
system which tends to be very compact and presumably leads to strong clusterization
in the real system followed by loss of atoms from the trap.

For fermions (not shown), the Pauli principle 
and the subsequent need to occupy higher orbitals means that the system will 
in general be larger for the same scattering length and will not show a decreasing
behaviour as $N$ is increased, but rather a slight increase. This can be 
seen from the energetics of the ground state in (\ref{em4}) which is 
dominated by the positive term containing the oscillator frequency for 
large $N$. This implies that the radius will stay large, constrained 
solely by the external trap.

We notice that the 
dimensional effect on the radius is quite clear when comparing to previous 
studies of bosonic systems in two and three dimensions \cite{arms11}. 
In the current one dimensional study, we see a non-monotonic behaviour in the 
radii as function of $N$ which can also be seen in two dimensions but which is 
almost completely absent in the corresponding three-dimensional system. The
scattering lengths for which the maximum in the size occurs are roughly those
where the three-body energy comes out positive. This implies an increased 
radial size, although still restricted by the external trapping potential. 
For large $N$, however, the bosons will always become negative in energy
as the shift term dominates in (\ref{em3}) and the radius goes down again. 
The fact that lower dimensions show a more pronounced non-monotonicity of the
size can then be traced to the fact that the positive energy contribution 
from zero-point motion grows with dimension and washes out the behaviour. 
For fermions, as discussed above, the energy is positive from the start
and the radius stays large. However, the degeneracies allowed in higher
dimensions adds extra ingredients and shell-structure to both energy and
to the radial size.

\subsection{Three boson energies}
We now consider the case of three bosons within our harmonic 
approximation scheme in order to make a comparison to the 
results obtained by MacGuire \cite{macguire1964} in the absence
of external confinement. The exact bound state energy, $E_N$, of $N$ bosons interacting
via attractive pairwise zero-range interactions in a 
homogeneous one-dimensional space can be expressed in term
of the two-body energy, $E_2$, as
\begin{equation}
E_{N}=\frac{1}{6}N\left(N^2-1\right)E_2.
\end{equation}
For the case of three particles this becomes $E_3=4E_2$. Since we 
have an external trap, we do not expect to reproduce this result.
However, in the limit $a\to 0$ where $E_2\to -\infty$ the external
trap should become negligible and a comparison can be made.

\begin{figure}
\includegraphics[width=0.5\textwidth]{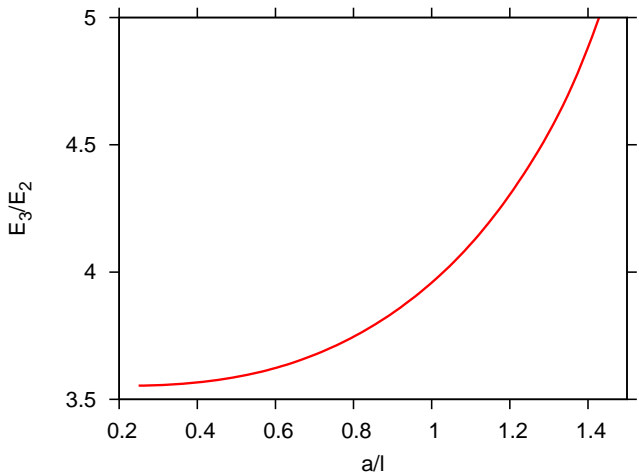}
\caption{The ratio $E_3/E_2$ as a function of scattering length for bosons.  
We plot only the cases where both $E_3$ and $E_2$ are negative.}
\label{fr3}
\end{figure}

\begin{figure}
\includegraphics[width=0.5\textwidth]{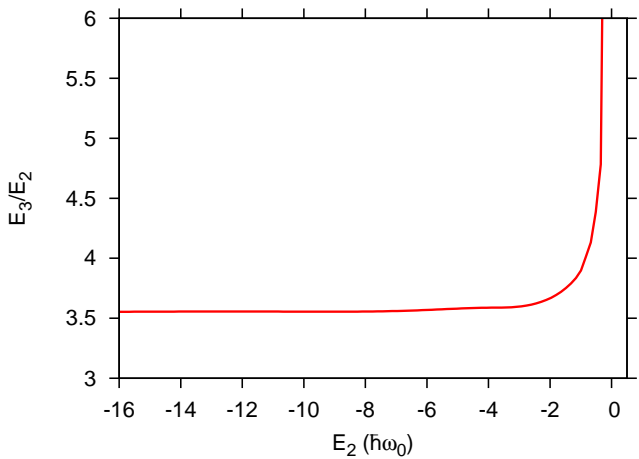}
\caption{The ratio $E_3/E_2$ as a function of $E_2$ for bosons.  
We plot only the cases where both $E_3$ and $E_2$ are negative.}
\label{fr4}
\end{figure}

We plot the ratio $E_3/E_2$ as a function of 
scattering length in figure \ref{fr3} and as a function of $E_2$ in figure \ref{fr4}.  
The plot ends at $a/l\sim 0.25$ at which point it becomes numerically 
very challenging to compute the wave function based on the exact
solution of Busch {\it et al.} \cite{busc98}. However, beyond that we
can use the exact solution as discussed below.

One can see in figure \ref{fr3} as the scattering 
length decreases (more clearly seen as $E_2\to -\infty$ in figure \ref{fr4}), 
that the ratio approach a limit.  Using (\ref{em3}) 
and (\ref{em8}) in the limit where $a\to 0$ and $\omega_r\to\omega_{in}$ (see (\ref{em5})), we get that
\begin{equation}
\frac{E_3}{E_2}=3+\left(\sqrt{\frac{3}{2}}-\frac{3}{2}\right)\frac{\omega_{in}}{E_2}.
\end{equation}
Relating $\omega_{in}$ to $\langle x^2\rangle$ through (\ref{em7}), we obtain
\begin{equation}
\frac{E_3}{E_2}=3+\left(\sqrt{\frac{3}{2}}-\frac{3}{2}\right)\frac{\hbar^2}{2\mu E_2\langle x^2\rangle}.
\end{equation}
We can now use the exact wave function for a delta function potential in one 
dimension, $\psi(x)=Ae^{-\kappa x}$ where $\kappa=\sqrt{-2\mu E_2/\hbar^2}$, 
to obtain $\hbar^2/(2\mu E_2\langle x^2\rangle)=2$. Our limit thus becomes
$E_3/E_2\approx 3.55$. This is also the number we find numerically for the
smallest value of $a$ that we could access. Comparing to the exact
value of MacGuire, $E_3/E_2=4$, this implies that the energetics of
our model is accurate to around 10\% for small scattering 
lengths. The wave function in the harmonic model is a 
gaussian which in the limit $a\to 0$ will tend to a delta
function, similar to the exact solution, so we also expect the
harmonic model to provide an accurate structural description in 
this limit.

On the other hand, when $a$ becomes very large (the unitarity limit)
the trap plays an important role. Looking back at the basic 
Hamiltonian in (\ref{em1}), we see that when $a\to\infty$, the 
interaction term tends to zero and we should be dominated by 
the trap only. However, the exact result does show that there
is a shift of the energy in this limit so that $E_2=\hbar\omega/2$
for $a\to\infty$ (the non-interacting system has twice the 
energy since there is zero-point motion from both particles).
Within our model we have $\omega_r\to\omega_0$ and $\langle x^2\rangle\to l^2/2$ 
when $a\to\infty$
as the external field provides the only length scale left in the problem. 
We have
\begin{equation}
\frac{E_3}{E_2}=3-\frac{\hbar\omega_0}{2E_2},
\end{equation}
and in the large $a$ limit, $E_2\to\hbar\omega_0/2$, so we obtain a limiting 
value of $E_3\to 2$.  At $a/l=100$, we numerically obtain 1.98 for this ratio so 
we reproduce this limit. The wave function becomes essentially exact in this
limit, and we thus expect the structure to be well reproduces also. 

In the limit of large $N$, we can also estimate the scalings and compare to the 
exact results when $a\to 0$. From (\ref{em3}) we have that the bosons scale
as $E_N\sim N(N-1)E_2$, which means that we have an underbinding by a factor 
of $N+1$. Fermions in (\ref{em4}) are different since here we have $E_N\sim N^{5/2}E_2$, 
i.e. the underbinding is only by a factor of about $\sqrt{N}$. Notice, however, 
that in the opposite limit of $a\to \infty$, the dominant term in (\ref{em3})
and (\ref{em4}) change as $V_S\to 0$ and $\omega_r\to\omega_0$. This is an 
essentially non-interacting situation and we have the intuitively obvious
result $E_N\sim N\hbar\omega_0/2$, which is just the scaling dictated by
the trap.  For values of $a$ away from these two limits, the scaling is 
an interplay of both terms and we get a more complicated behaviour. We note
that one could also turn these arguments upside down and use the exact 
results for $N$ particles in the $a\to 0$ limit to fit the parameters
of the model, for instance by using $E_3=4E_2$ instead of either the 
frequence of radius condition we impose on the two-body problem. Overall,
we expect this to provide only minor quantitative changes in the harmonic
model and its predictions.

\subsection{One-body density}
\begin{figure}
\includegraphics[width=0.5\textwidth]{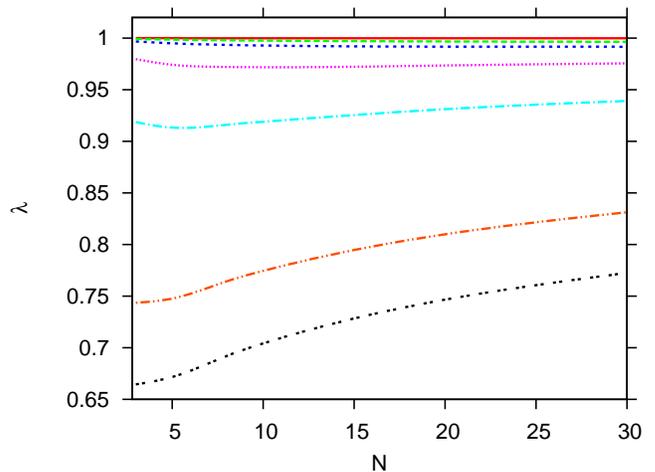}
\caption{  The value of $\lambda$ at several values of the 1D scattering length.  
From bottom to top they are 0.4, 0.5, 1, 2, 5, 10, and 100.}
\label{fr6}
\end{figure}

The one body density matrix
\begin{equation}
\rho(x_1,x_1')=N\int\psi^*(x_1,\dots,x_N)\psi(x_{1}^{'},\dots,x_N)dx_2\dots dx_N, 
\end{equation}
is the most basic correlation function and is the starting point of statistical 
property calculations. It contains information about the mean-field nature of 
the wave function for a bosonic system.  
It is of interest in itself, but we will focus on its 
largest eigenvalue, $\lambda$, which directly measures how much a given state 
has the structure of a coherent state. Figure \ref{fr6} shows $\lambda$ as a 
function of the number of bosons for several scattering lengths.  We mainly 
see an increase with particle number, though there are slight decreases for small 
particle numbers and large scattering lengths.  In contrast with our previous 
results in \cite{arms11} for higher dimensions, only scattering lengths of order 
unity or smaller show a $\lambda$ much different than one, showing that the 
persistence of the mean-field structure provided by the external potential is 
quite strong in one dimension implying that the mean-field approximation is 
very good. The difference to higher dimensions arise from the degeneracies
and from the reduction implied by the fact that one can factorize
$\lambda$, so that for instance $\lambda_{3D}=\lambda^3$ (see \cite{arms11}).

\subsection{Thermodynamics}
The energy spectrum can be obtained by considering center-of-mass excitations 
in the system and also the internal excitations characterized by the 
frequency in (\ref{em5}).
To address the thermodynamics of the systems, one can proceed as 
described in detail in \cite{arms12}.  One naturally starts with the partition function
\begin{equation}
Z(T)=\sum_i g_i\exp[-E_i/(k_BT)],
\end{equation}
where $E_i$ is the energy of the $i$th state, $g_i$ is the degeneracy of that 
state, and $k_B$ is Boltzmann's constant.  
1D systems are simpler since there is no degeneracy in the energy levels in contrast  
to higher dimensions.  With the sequence of energies and single-particle degeneracies, 
thermodynamic quantities can be calculated.  Once the spectrum and degeneracies 
of our systems have been worked out, 
we can calculate basic thermodynamic quantities in the canonical 
ensemble.  In 1D, the number of states grows much more slowly than in higher 
dimensions, though one does have to climb somewhat high in the spectrum due to 
the lack of degeneracy.  In fact, in 1D there is no difference between bosons 
and fermions in the sequence of the degeneracies of the $N$-body excited states. This 
comes from the fact that all the Pauli principle implies for fermions is that the 
ground state has them occupying higher orbit which increases the ground state
energy in comparison to bosons which can all go into the lowest orbital. Excitations
on top of these ground states will now cost the same amount of energy and they will all
have the same degeneracy since it is simply a matter of distribution of a number of
particles in levels to obtain a specific total energy cost (the energy of the excitation
above the ground state).

\begin{figure}
\includegraphics[width=0.5\textwidth]{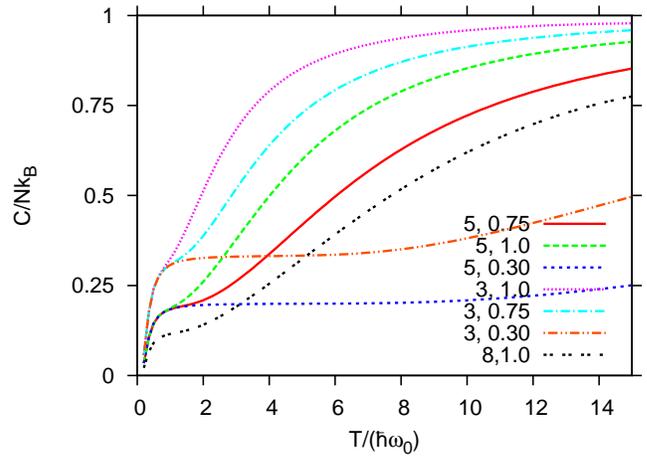}
\caption{The heat capacity per particle at several boson numbers and 
scattering lengths, identified as $N, a$, as a function of temperature.}
\label{fr7}
\end{figure}

This latter fact implies that 
some thermodynamic quantities such as the entropy and heat capacity are identical 
for bosons and fermions at the same scattering length.  Working in the 
canonical ensemble, in figure \ref{fr7} we show the heat capacity $C$, which is 
the same for bosons and fermions.  The heat capacity is related to the partition 
function by
\begin{eqnarray}
C=\frac{\partial E}{\partial T},\\
E=-k_BT^2\frac{\partial \log Z}{\partial T}.
\end{eqnarray}
For all particle numbers, we see an initial increase, which then slows 
down at higher temperatures.  This initial increase is the saturation 
of the center of mass mode.  The temperature, where the slope changes, 
depends on particle number since more particles present in the system implies that 
the center of mass mode comprises a smaller share of the overall excitation energy.  
At the larger scattering lengths, the heat capacity then gradually increases 
towards the equipartion limit value of one.  At small scattering lengths, 
the large size of the excitation frequency sharply delays the approach to 
the equipartition limit.  For the same scattering length, more particles 
mean a slower approach to the high temperature limit.  This is clear from 
(\ref{em5}), which shows that the excitation frequency increases with 
particle number.  

\section{Summary and outlook}
We have presented results for an $N$-body system of particles
moving in one dimension and interacting via pairwise 
zero-range interactions under the influence of an external harmonic oscillator 
trap. The Hamiltonian is solved analytically within the harmonic approximation 
where the parameters of the model are adjusted to reproduce properties of the 
exact solution of the corresponding two-body problem. This is along the 
lines of the standard approach to many-body problems in many fields of 
physics. 

The ground state energies for both bosonic and fermionic systems were 
determined and we discussed the scaling of these quantities with the particle 
number. We also obtained the radial size of the bosonic systems and compared
to previous findings in two- and three-dimensional setups. A careful study 
of the energy of three bosons in the harmonic model reveals that in the 
limit of large two-body binding energy where the external trap can be 
neglected, the harmonic model results are about 10\% below the exact 
result abtained by MacGuire \cite{macguire1964}, while in the opposite
limit of negligible two-body binding energy (large scattering length)
the model becomes essentially exact. For larger particle numbers, the 
harmonic approximation tends to underbind the system compared to the exact
results when the two-body system is strongly bound.

Additionally, we studied the one-body density matrix. The main goal was
the computation of its largest eigenvalue to characterize the degree
of coherence in the system. The results turn out to be similar to those
obtained in two and three dimensions, except that the one-dimensional 
coherence turns out to be more robust except for large two-body binding
energies. Lastly, we studied the thermodynamics of the one-dimensional 
system. The specific heat was found to behave in a manner that is also
quite similar to higher dimensions. However, it is worth noticing that 
within the harmonic model, the spectrum of excitation for bosons and 
fermions turns out to be exactly the same. Due to the lack of degeneracy
of the levels in one dimensions, the only difference is therefore found 
in the ground state energy.

The one-dimensional setup studied here provides a very valuable benchmark for
the harmonic models that have been used in higher dimensions in different 
contexts \cite{calogero1971,zaluska2000,yan2003,gajda2006}. Another good 
test ground for the harmonic models is systems that have long-range 
interaction. In particular, systems of dipolar molecules in one-dimensional 
tubes have been studied recently due to their interesting few- and 
many-body structure \cite{arguelles2007,kollath2008,chang2009,huang2009,wunsch2011,dalmonte2011,zinner2011,fellows2011,knap2012,leche2012}.
The dipole-dipole interaction in two different one-dimensional tubes 
is typically characterized by having a pocket which becomes deep 
when the dipolar interaction is strong \cite{zinner2011}. This implies 
that a harmonic approximation is a good starting point. This approach 
has been studied in the case of two-dimensional layers with 
dipolar particles \cite{wang2006,jeremy2010,zinner2011d,jeremy2011} and 
compares very well to exact results \cite{artem2011a,artem2011b,artem2012} 
away from the weak-coupling limit. The application of the harmonic 
approximation to one-dimensional arrays containing dipolar particles is 
a topic for future work.

Another interesting direction in which the harmonic model can be used
is for the study of higher-order terms in the interaction such as 
those coming from effective-range corrections to the model of 
Busch {\it et al.} \cite{busc98,zinner2012}. Recent experiments have 
shown a potential need for such corrections in both two- \cite{baur2012}
and three-dimensional cold atomic gas systems \cite{ohara2012}. A 
possible way to account for such corrections would be to include quartic
interaction terms an then carefully fit the additional parameter to few-body 
properties. Within the harmonic model, the quartic interactions can be 
computed quite straightforwardly through gaussian integrals. This 
could then be compared to mean-field results that include effective-range
corrections in the limit of large particle numbers \cite{fu2003,zinner2009a,zinner2009b,thoger2009}.

\end{document}